\documentclass[aps,pra,showpacs,groupedaddress,superscriptaddress,twocolumn]{revtex4}
\usepackage{makeidx}
\usepackage{amssymb}
\usepackage{amsmath}
\usepackage[dvips]{graphics,color}
\usepackage{graphicx}
\usepackage{latexsym}

\newcommand{\be}{\begin{equation}}
\newcommand{\ee}{\end{equation}}
\newcommand{\br}{\begin{eqnarray}}
\newcommand{\er}{\end{eqnarray}}

\newcommand{\bd}{\begin{displaymath}}
\newcommand{\ed}{\end{displaymath}}

\newcommand{\bfig}{\begin{figure}}
\newcommand{\efig}{\end{figure}}
\input{tcilatex}

\begin{document}

\title{Inclusion of non-idealities in the continuous photodetection model}
\author{A. V. Dodonov}
\email{adodonov@df.ufscar.br}
\author{S. S. Mizrahi}
\email{salomon@df.ufscar.br}
\affiliation{Departamento de F\'{\i}sica, CCET, Universidade Federal de S\~{a}o Carlos,
Via Washington Luiz km 235, 13565-905, S\~ao Carlos, S\~ao Paulo, Brazil}
\author{V. V. Dodonov}
\email{vdodonov@fis.unb.br}
\affiliation{Instituto de F\'{\i}sica, Universidade de Bras\'{\i}lia, PO Box 04455,
70910-900, Bras\'{\i}lia, Distrito Federal, Brazil}
\date{\today }

\begin{abstract}
Some non-ideal effects as non-unit quantum efficiency, dark counts, dead
time and cavity losses that occur in experiments are incorporated within the
continuous photodetection model by using the analytical quantum trajectories
approach. We show that in standard photocounting experiments the validity of
the model can be verified, and the formal expression for the quantum jump
superoperator can also be checked.
\end{abstract}

\maketitle

\section{Introduction}

The continuous photodetection model (CPM) was proposed in the early 1980's
in order to treat quantum optics situations in which a weak electromagnetic
field enclosed in a cavity is continuously measured through the
photocounting approach \cite{SD}. The theory has received considerable
attention in the following years due to its new microscopic interpretation
of the photodetection process \cite{ueda1,ueda2,ueda4,ueda5,ueda6}, relation
to the quantum trajectories approach \cite{QT1,QT11,QT2,QT3,QT4,QT5} and
several proposals for applications. Among them we find studies of
photocounts statistics in diverse systems \cite{a2,a5,a8,a10,a17}, quantum
non-demolition measurements \cite{a15,ueda3,walls}, implementation of
measurement schemes \cite{a1,a4,a14}, quantum state preparation \cite%
{a9,a11,a12,a13,a16}, quantum control via photodetection \cite{a6,a7}, and
quantum computation \cite{a3}.

CPM is extensively discussed in the literature \cite%
{ueda2,a15,SD1,SD2,OMD-JOB}, so we shall mention only its main properties.
The model, also referred as a theory, describes the field state evolution
during the photodetection process in a closed cavity and is formulated in
terms of two fundamental \textit{operations}, assumed to represent the
\textit{only} events taking place at each infinitesimal time interval. (1)
The one-count operation, represented by the \textit{Quantum Jump
Superoperator} (QJS), describes the detector's action on the field upon a
single count, and the trace calculation over the QJS gives the probability
per unit time for occurrence of a detection. (2) The \textit{no-count}
operation describes the field non-unitary evolution in absence of counts.

If one sets the formal expressions for these operations, all possible
outcomes of a photocounting experiment can be predicted. For instance, the
photocounts \cite{SD,ueda1,ueda2} and the waiting time \cite%
{waiting1,waiting2,waiting3,waiting4} statistics are among the most common
quantities to be studied both theoretically and experimentally. Moreover,
CPM conferred a new step in photodetection theories by allowing to determine
the field state after an arbitrary sequence of measurements, thus creating
the possibility of controlling the field properties in real time experiments
\cite{a7,a10,a17}.

Actually, the QJS is the main formal ingredient within the theory, since it
also dictates the form of the no-count superoperator \cite{SD}. Two
different models for the QJS were proposed \emph{ad hoc}. The first one was
proposed by Srinivas and Davies \cite{SD}, the \textit{SD-model}, as
\begin{equation}
\hat{J}\rho =\lambda \hat{a}\rho \hat{a}^{\dagger },  \label{01}
\end{equation}%
where $\rho $ is the field density matrix, $\hat{a}$ and $\hat{a}^{\dagger }$
are the usual bosonic ladder operators and $\lambda $ is roughly the
detector's ideal counting rate \cite{SD,DMD-JOB05}. From the very beginning
the authors \cite{SD} denounced the presence of some inconsistences when the
QJS (\ref{01}) is employed for describing a real photodetection process,
this point was also appointed in \cite{DMD-JOB05}. Nevertheless, this QJS is
widely used in the literature \cite%
{ueda2,ueda4,ueda6,a2,a4,a6,a7,a9,a10,a12,a13,a16}.

The other proposal \cite{OMD-JOB,benaryeh} assumes for the QJS an expression
written in terms of the ladder operators $\hat{E}_{-}=\left( \hat{a}%
^{\dagger }\hat{a}+1\right) ^{-1/2}\hat{a}$ and $\hat{E}_{+}=\hat{E}%
_{-}^{\dagger }$ (also known as \emph{exponential phase operators} \cite%
{p1,p2,Vourd92,p3,p4})
\begin{equation}
\hat{J}\rho =\lambda \hat{E}_{-}\rho \hat{E}_{+}.  \label{02}
\end{equation}%
In \cite{DMD-JOB05} we called \emph{E-model} such a choice, to differentiate
from the SD QJS (\ref{01}). Besides eliminating the inconsistencies within
the SD-model, the use of\ the E-model leads to different qualitative and
quantitative predictions for several observable quantities. By an analysis
of a microscopic model for the detector, it was recently shown that the
QJS's (\ref{01}) and (\ref{02}) are particular cases of a general
time-dependent \emph{transition superoperator}, each one occurring in a
particular regime of the detector experimental parameters \cite{QJS,EQJS}.
Moreover, it was pointed out that by manipulating certain detector's
accessible parameters one could engineer the form of the QJS, thus changing
the dynamics of the photodetection, as well as the field state after a
sequence of measurements.

A way to check the validity of the CPM and to decide which QJS better
describes the phenomenon in practice can be accomplished through photocount
experiments in a high finesse cavity by comparing the results to the
theoretical predictions. However, real detectors and cavities are far from
ideal. So our first goal is to include into the CPM the effects of
non-ideality, such as quantum efficiency (QE), dark counts, detector's
dead-time and cavity damping. Our second goal is to call attention to the
fact that standard photodetection measurements could verify which of the QJS
models actually prevails experimentally.

The plan of the paper is as follows. In Sec. \ref{s2}, we present a simple
model, which enables us to include the effects of non-ideality -- QE and
dark counts -- into the CPM using the quantum trajectories approach. Then we
calculate the main quantities characterizing the photodetection process --
the photocounting and waiting time distributions. In subsection \ref{suba}
we do this using the QJS (\ref{01}), and in subsection \ref{subb} we repeat
the same procedure for E-model. In Sec. \ref{s3}, we analyze the behavior of
the lower moments of the above distributions in realistic situations and
point out how one could decide about a QJS from experimental data. Sec. \ref%
{s4} contains conclusions. In the appendix \ref{apendix1} we treat the
effects of dead-time and cavity damping: we show that 1) cavity losses are
not significant compared to non-unit QE effect, 2) the dead-time effect
leads to mathematical inconsistences in SD-model, yet it is free of them in
E-model, being however quite small compared to QE effect. Appendix \ref%
{apendix2} contains some mathematical details concerning evaluation of
quantities of interest for different quantum states.

\section{Models of non-ideal photodetectors}

\label{s2}

\subsection{SD-model}

\label{suba}

We consider a free electromagnetic mono-modal field of frequency $\omega $,
enclosed in an ideal cavity together with a photodetector (in the appendix %
\ref{apendix1} we show that the cavity damping is not crucial if the
detector has non-unit QE). The \emph{unconditioned time evolution} (UTE) of
the field in the presence of the detector, i.e. the evolution when the
detector is turned on but the outcomes of the measurements are disregarded
(not registered), is described by the master equation \cite{a15,OMD-JOB,a4}
\begin{equation}
\dot{\rho}=-i\omega \left( \hat{n}\rho -\rho \hat{n}\right) -\frac{\lambda }{%
2}\left( \hat{n}\rho +\rho \hat{n}-2\hat{A}\rho \right) ,  \label{nc}
\end{equation}%
where $\hat{A}\rho \equiv \hat{a}\rho \hat{a}^{\dagger }$ is a superoperator
and $\hat{n}=\hat{a}^{\dagger }\hat{a}$ is the number operator. The first
term stands for the free field evolution while the second describes the
effect of the detector on the field due to their mutual interaction. The
parameter $\lambda $ is the field-detector coupling constant, roughly equal
to the ideal counting rate \cite{QJS,EQJS}.

To describe photocounting with QE $\eta $ and finite dark counts rate $%
\lambda d$ ($d$ is the ratio between the dark counts rate and the ideal
photon counting rate), we assume the following expression for the QJS (c.f.
the expression resulting from the microscopic model in \cite{EQJS})
\begin{equation}
\hat{J}\rho =\lambda \left( \eta \hat{A}+d\right) \rho .  \label{qjs}
\end{equation}%
It describes the action of the detector on the field upon a photodetection,
and its trace gives the probability per unit time of the click. Actually,
the microscopic model \cite{QJS} suggests that (\ref{qjs}) has a diagonal
form in the Fock basis, but this will not be important here, since we shall
be interested only in diagonal elements. The first term within the
parenthesis describes the absorption of a photon from the field with
probability per unit time $\mathrm{Tr}\left[ \eta \lambda \hat{A}\rho \right]
=\eta \lambda \bar{n}$, where $\bar{n}$ is the field mean photon number --
this means that the detector `sees' all the photons. The second term
describes the occurrence of a dark count with field-independent probability
density $\lambda d$, and this event by itself does not modify the field
state (the field state after a single dark count is $\lambda \rho d/\mathrm{%
Tr}\left( \lambda \rho d \right) =\rho $). However, when both terms are
present, the field state upon a detector's click becomes a mixture of both
outcomes.

From the quantum trajectories approach and CPM \cite{SD,ueda1,QT1}, all the
quantities related to photodetection can be calculated provided the
complementary no-count superoperator $\hat{S}_{t}$ is known ($\hat{S}_{t}$
describes the action of the detector on the field during the time interval $%
t $ without registered counts). Acting $\hat{S}_{t}$ on the initial field
state $\rho _{0}$, the no-count state $\rho _{S}\equiv \hat{S}_{t}\rho _{0}$
obeys Eq. (\ref{nc}) when one subtracts the term (\ref{qjs}) on the RHS (see
\cite{QT1,QT11}). Moreover, as we are interested in calculating
probabilities, we shall disregard phase factors $\exp (\pm i\omega \hat{n}t)$%
, since they are canceled in any trace evaluation. So the evolution equation
of $\rho _{S}$ is
\begin{equation}
\dot{\rho}_{S}=-\frac{\lambda }{2}\left( \hat{n}\rho _{S}+\rho _{S}\hat{n}%
\right) +\lambda q\hat{A}\rho _{S}-\lambda d\rho _{S},\quad q\equiv 1-\eta .
\label{sol}
\end{equation}%
Setting the transformation
\begin{equation}
\rho _{S}=e^{-d\lambda t}\hat{U}_{t}\rho _{1},\qquad \hat{U}_{t}\rho
=e^{-\lambda t\hat{n}/2}\rho e^{-\lambda t\hat{n}/2}
\end{equation}%
in Eq. (\ref{sol}) we obtain a simple equation for $\rho _{1}$
\begin{equation}
\dot{\rho}_{1}=\lambda qe^{-\lambda t}\hat{A}\rho _{1},
\end{equation}%
whose solution is%
\begin{equation}
\rho _{1}=\sum_{l=0}^{\infty }\frac{\left( q\phi _{t}\right) ^{l}}{l!}\hat{a}%
^{l}\rho _{0}\left( \hat{a}^{\dagger }\right) ^{l}\equiv \exp {(q\phi _{t}%
\hat{A})}\rho _{0},  \label{wdv}
\end{equation}%
where%
\begin{equation}
\phi _{t}=1-e^{-\lambda t}.  \label{def-phit}
\end{equation}

Thus the no-count superoperator is%
\begin{equation}
\hat{S}_{t}\rho _{0}=e^{-d\lambda t}\hat{U}_{t}(e^{q\phi _{t}\hat{A}}\rho
_{0}).  \label{st}
\end{equation}%
The field UTE superoperator $\hat{T}_{t}$, defined as the solution to Eq. (%
\ref{nc}), is naturally given by setting $d=\eta =0$ in Eqs. (\ref{sol}) and
(\ref{st}), i.e.%
\begin{equation}
\hat{T}_{t}=\hat{U}_{t}(e^{\phi _{t}\hat{A}}\rho _{0}).
\end{equation}%
We introduced in Eq. (\ref{wdv}) a compact notation for the infinite sum in
terms of the exponential superoperator. We can deal with such superoperators
as they were common operators, provided we use the `commutation relations'
\begin{equation}
\hat{A}\hat{U}_{t}=e^{-\lambda t}\hat{U}_{t}\hat{A},\qquad e^{y\hat{A}}\hat{U%
}_{t}=\hat{U}_{t}\exp (ye^{-\lambda t}\hat{A}),  \label{exx}
\end{equation}%
obtained by expanding the superoperators in series.

Now we can calculate the $m$-counts superoperator $\hat{N}_{t}(m)$, that
describes the field state after $m$ registered counts (whatever real or dark
ones) in the time interval $(0,t)$, and whose trace gives the probability
for this event. It reads
\begin{equation}
\hat{N}_{t}(m)\rho =\int \cdots \int \hat{h}\rho ,  \label{n-nodead}
\end{equation}%
where the integrals are evaluated over all the time intervals between the
counts%
\begin{equation}
\int \cdots \int \equiv \int_{0}^{t}dt_{m}\int_{0}^{t_{m}}dt_{m-1}\cdots
\int_{0}^{t_{2}}dt_{1}
\end{equation}%
and the \emph{conditioned density operator} is\emph{\ }%
\begin{equation}
\hat{h}\rho \equiv \hat{S}_{t-t_{m}}\hat{J}\hat{S}_{t_{m}-t_{m-1}}\hat{J}%
\cdots \hat{J}\hat{S}_{t_{1}}\rho .  \label{rho_c}
\end{equation}%
Expanding the QJS (\ref{qjs}) in Eq. (\ref{rho_c}) in terms of $\eta \hat{A}
$ and $d$, one obtains a finite sum whose first term, proportional to $d^0$,
describes the detection of $m$ photons:
\begin{eqnarray}
\hat{h}^{(0)} &=&\left( \lambda \eta \right) ^{m}\hat{S}_{t-t_{m}}\hat{A}%
\cdots \hat{A}\hat{S}_{t_{1}}  \label{n01} \\
&=&\left( \lambda \eta \right) ^{m}e^{-\lambda \left( t_{1}+t_{2}+\cdots
+t_{m}\right) }\hat{S}_{t}\hat{A}^{m}.  \notag
\end{eqnarray}%
After integrating (\ref{n01}) we obtain the first term in (\ref{n-nodead}),
describing the field state after the loss by absorption of $m$ photons,
\begin{equation}
\hat{n}_{t}(m)\equiv \int \cdots \int \hat{h}^{(0)}=\hat{S}_{t}\frac{\left(
\eta \phi _{t}\hat{A}\right) ^{m}}{m!}.  \label{n1}
\end{equation}%
Calculating in a similar way the contribution of the terms with higher
powers in $d$ we arrive at the formula
\begin{equation}
\hat{N}_{t}(m)=\sum_{k=0}^{m}\frac{\left( d\lambda t\right) ^{k}}{k!}\hat{n}%
_{t}(m-k)=\hat{S}_{t}\frac{(d\lambda t+\eta \phi _{t}\hat{A})^{m}}{m!}.
\label{n12}
\end{equation}%
One can easily verify that the $m$-counts superoperators (\ref{n12}) satisfy
identically the fundamental relation \cite{SD,QT1}%
\begin{equation}
\sum_{m=0}^{\infty }\hat{N}_{t}(m)=\hat{T}_{t}.  \label{tt1}
\end{equation}

The factorial moments of the photocounts distribution are easily evaluated
as
\begin{gather}
\overline{m\cdots (m-l)}_{t}=\sum_{m=0}^{\infty }m\cdots (m-l)\mathrm{Tr}[%
\hat{N}_{t}(m)\rho ]  \notag \\
=\mathrm{Tr}[\hat{S}_{t}(d\lambda t+\eta \phi _{t}\hat{A})^{l+1}\exp
(d\lambda t+\eta \phi _{t}\hat{A})\rho ]  \notag \\
=\mathrm{Tr}[\hat{U}_{t}(d\lambda t+\eta \phi _{t}\hat{A})^{l+1}e^{\phi _{t}%
\hat{A}}\rho ].
\end{gather}%
Thus we need to calculate the expression%
\begin{eqnarray}
\Phi _{k}\left( b,x\right) &\equiv &\mathrm{Tr}\left[ \hat{U}_{b}e^{x\hat{A}}%
\hat{A}^{k}\rho \right]  \notag \\
&=&\sum_{n,l=0}^{\infty }\frac{\left( n+l+k\right) !}{n!l!}e^{-\lambda
bn}x^{l}\rho _{n+l+k}  \notag \\
&=&\sum_{n=k}^{\infty }\rho _{n}\frac{n!}{(n-k)!}\left( x+e^{-\lambda
b}\right) ^{n-k},  \label{lordi}
\end{eqnarray}%
where $\rho _{n}=\langle n|\rho |n\rangle$. Evaluating
\begin{equation}
\Phi _{k}\left( t,\phi _{t}\right) =\sum_{n=0}^{\infty }\rho _{n}\frac{n!}{%
(n-k)!}
\end{equation}%
(see Eq. (\ref{def-phit}) for the expression of $\phi_t$) we obtain general
expressions for the lower factorial moments%
\begin{equation}  \label{msd}
\bar{m}_{t}=d\lambda t+\eta \bar{n}\phi _{t}
\end{equation}%
\begin{equation}
\overline{m(m-1)}_{t}=\left( d\lambda t\right) ^{2}+2\eta \bar{n}d\lambda
t\phi _{t}+(\eta \phi _{t})^{2}\overline{n(n-1)},
\end{equation}%
where $\bar{n}$ and $\overline{n(n-1)}$ are the factorial moments of the
initial density operator.

Another measurable quantity we consider here is the \emph{waiting time
distribution}. It describes the probability density for registering two
consecutive clicks separated by the time interval $\tau $, under the
condition that the first one occurred at time $t$. Its non-normalized form is%
\begin{equation}
\mathrm{W}_{t}(\tau )=\mathrm{Tr}\left[ \hat{J}\hat{S}_{\tau }\hat{J}\hat{T}%
_{t}\rho \right] ,  \label{www}
\end{equation}%
and the mean waiting time is%
\begin{equation}
\bar{\tau}=\mathcal{N}^{-1}\int_{0}^{T}d\tau \mathrm{W}_{t}\left( \tau
\right) \tau ,\quad \mathcal{N}=\int_{0}^{T}d\tau \mathrm{W}_{t}\left( \tau
\right) ,
\end{equation}%
where $T$ is the time interval during which one evaluates the averaging in
experiments. As will be shown in section \ref{s3}, $T$ is an important
parameter due to the presence of dark counts. After straightforward
manipulations, using the `commutation relations' (\ref{exx}), we obtain
\begin{eqnarray*}
{\mathrm{W}}_{t}(\tau ) &=&e^{-d\lambda \tau }\left[ \eta ^{2}e^{-\lambda
\left( 2t+\tau \right) }\Phi _{2}^{W}\right. \\
&&\left. +\eta de^{-\lambda t}(1+e^{-\lambda \tau })\Phi _{1}^{W}+d^{2}\Phi
_{0}^{W}\right] ,
\end{eqnarray*}%
where
\begin{equation}
\Phi _{k}^{W}= \Phi _{k}\left[ t+\tau ,1-e^{-\lambda t}\left( \eta +(1-\eta
)e^{-\lambda \tau }\right) \right] .  \label{fkw}
\end{equation}

In the appendix \ref{apendix1} we consider the dead-time effect and show
that it cannot be consistently incorporated into SD-model, because the QJS (%
\ref{qjs}) is an unbounded superoperator and the resulting counting
probability is non-normalizable. This is just one more mathematical
inconsistency \cite{DMD-JOB05} of the SD-model. In the appendix \ref%
{apendix2} we evaluate the expression (\ref{lordi}) for three kinds of
states: coherent, number and thermal.

\subsection{E-model}

\label{subb}

We now repeat the same procedures for E-model in which the QJS is
\begin{equation}
\hat{J}\rho =\lambda \left( \eta \hat{\varepsilon}+d\right) \rho ,
\end{equation}%
where $\hat{\varepsilon}\rho \equiv \hat{E}_{-}\rho \hat{E}_{+}$. The
probability per unit time for detecting a photon is $\eta \lambda (1-p_{0})$%
, where $p_{0}=\langle 0|\rho |0\rangle $, so the detector `sees' whether
there is any photon in the cavity. In principle, the parameter $\lambda $ is
different from the one in SD-model, but in the context of this paper it will
be always clear which one we are dealing with. The field UTE is described by
an equation similar to Eq. (\ref{nc}), obtained by doing the substitution $%
\left\{ \hat{a},\hat{a}^{\dagger }\right\} \rightarrow \{\hat{E}_{-},\hat{E}%
_{+}\}$ in the non-unitary evolution (second term on the RHS). So the
no-count state $\rho _{S}$ obeys the equation
\begin{equation}
\dot{\rho}_{S}=-\frac{\lambda }{2}\left( \hat{\Lambda}\rho _{S}+\rho _{S}%
\hat{\Lambda}\right) +\lambda q\hat{\varepsilon}\rho _{S}-d\lambda \rho _{S},
\label{hgf}
\end{equation}%
[similar to Eq. (\ref{sol})] where $\hat{\Lambda}\equiv \hat{E}_{+}\hat{E}%
_{-}=1-\hat{\Lambda}_{0}$, $\hat{\Lambda}_{0}\equiv |0\rangle \langle 0|$.
Setting the transformation
\begin{equation}
\rho _{S}=e^{-d\lambda t}e^{-\lambda t\hat{\Lambda}/2}\rho _{1}e^{-\lambda t%
\hat{\Lambda}/2}
\end{equation}%
in Eq. (\ref{hgf}) and using the property $\exp (\alpha \hat{\Lambda})=\hat{%
\Lambda}_{0}+e^{\alpha }\hat{\Lambda},$ we obtain the differential equation
for $\rho _{1}$
\begin{eqnarray}
\dot{\rho}_{1} &=&\lambda qe^{-\lambda t}\left( \hat{\Lambda}_{0}+e^{\lambda
t/2}\hat{\Lambda}\right) \hat{E}_{-}\left( \hat{\Lambda}\rho _{1}\hat{\Lambda%
}\right)  \label{ggg} \\
&&\times \hat{E}_{+}\left( \hat{\Lambda}_{0}+e^{\lambda t/2}\hat{\Lambda}%
\right) .  \notag
\end{eqnarray}%
We solve this equation by projecting it onto orthogonal subspaces spanned by
projectors $\{\hat{\Lambda},\hat{\Lambda}_{0}\}$. Moreover, since at the end
we shall be interested only in calculating probabilities, we consider only
the diagonal part in Fock basis for quantities of interest, thus
disregarding the terms whose trace is null, such as $\hat{\Lambda}\rho \hat{%
\Lambda}_{0}$.

Multiplying Eq. (\ref{ggg}) by $\hat{\Lambda}$ on both sides we obtain
\begin{equation}
\frac{d}{dt}\left( \hat{\Lambda}\rho _{1}\hat{\Lambda}\right) =\lambda q\hat{%
\Lambda}\hat{E}_{-}\left( \hat{\Lambda}\rho _{1}\hat{\Lambda}\right) \hat{E}%
_{+}\hat{\Lambda},
\end{equation}%
whose solution is%
\begin{equation}
\hat{\Lambda}\rho _{1}\hat{\Lambda}=\hat{\Lambda}\left( e^{q\lambda t\hat{%
\varepsilon}}\rho _{0}\right) \hat{\Lambda}  \label{111}
\end{equation}%
and we note again that all the composite superoperators, such as $\exp (y%
\hat{\varepsilon}),$ are understood as power expansions. Now, multiplying (%
\ref{ggg}) by $\hat{\Lambda}_{0}$ on both sides and using the solution (\ref%
{111}) we get an equation for $\hat{\Lambda}_{0}\rho _{1}\hat{\Lambda}_{0}$
\begin{equation}
\frac{d}{dt}\left( \hat{\Lambda}_{0}\rho _{1}\hat{\Lambda}_{0}\right)
=\lambda q\hat{\Lambda}_{0}\left( e^{-\lambda t\left( 1-q\hat{\varepsilon}%
\right) }\hat{\varepsilon}\rho _{0}\right) \hat{\Lambda}_{0}
\end{equation}%
with solution%
\begin{equation}
\hat{\Lambda}_{0}\rho _{1}\hat{\Lambda}_{0}=\hat{\Lambda}_{0}\left[ \frac{1-q%
\hat{\varepsilon}\hat{R}_{t}}{1-q\hat{\varepsilon}}\rho_0\right] \hat{\Lambda%
}_{0},\quad \hat{R}_{t}\equiv e^{-\lambda t\left( 1-q\hat{\varepsilon}%
\right) }.
\end{equation}%
Thus the \emph{diagonal} form of the no-count superoperator, which we write
just in terms of the projector $\hat{\Lambda}_{0}$ and the unit operator, is%
\begin{equation}
\hat{S}_{t}\rho _{0}=e^{-d\lambda t}\left[ \hat{R}_{t}+\hat{\Lambda}_{0}%
\frac{1-\hat{R}_{t}}{1-q\hat{\varepsilon}}\hat{\Lambda}_{0}\right] \rho_0,
\label{ste1}
\end{equation}
where we use the notation $(\Lambda_0\hat{Q}\Lambda_0)\rho\equiv\Lambda_0(%
\hat{Q}\rho)\Lambda_0$.

Repeating the steps (\ref{n01}) -- (\ref{n12}), we obtain first the
conditioned density operator
\begin{equation}
\hat{h}^{(0)}\rho =e^{-d\lambda t}\left[ \hat{R}_{t}+\hat{\Lambda}_{0}\frac{%
\hat{R}_{t_{m}}-\hat{R}_{t}}{1-q\hat{\varepsilon}}\hat{\Lambda}_{0}\right]
(\lambda \eta \hat{\varepsilon})^{m}\rho .
\end{equation}%
After evaluating the time integrals as in (\ref{n1}) we get
\begin{eqnarray*}
\hat{n}_{t}(m) &=&e^{-d\lambda t}\left[ \left( 1-\hat{\Lambda}_{0}\frac{1}{%
1-q\hat{\varepsilon}}\hat{\Lambda}_{0}\right) \hat{R}_{t}\frac{\left(
\lambda t\eta \hat{\varepsilon}\right) ^{m}}{m!}\right. \\
&&\left. +\hat{\Lambda}_{0}\frac{\left( \lambda \eta \hat{\varepsilon}%
\right) ^{m}}{1-q\hat{\varepsilon}}\int_{0}^{t}dx\hat{R}_{x}\frac{x^{m-1}}{%
\left( m-1\right) !}\hat{\Lambda}_{0}\right]
\end{eqnarray*}%
for $m>0$ and $\hat{n}_{t}(0)=\hat{S}_{t}$. Finally, analogously to the
expression (\ref{n12}), we obtain the $m$-counts superoperator
\begin{eqnarray}
\hat{N}_{t}(m) &=&e^{-d\lambda t}\left\{ \left( 1-\hat{\Lambda}_{0}\frac{1}{%
1-q\hat{\varepsilon}}\hat{\Lambda}_{0}\right) \hat{R}_{t}\frac{(\hat{J}t)^{m}%
}{m!}\right.  \notag  \label{nde} \\
&&+\hat{\Lambda}_{0}\frac{1}{1-q\hat{\varepsilon}}\frac{\left( d\lambda
t\right) ^{m}}{m!}\hat{\Lambda}_{0} \\
&&\left. +\hat{\Lambda}_{0}\frac{\lambda \eta \hat{\varepsilon}}{1-q\hat{%
\varepsilon}}\int_{0}^{t}dx\hat{R}_{x}\frac{\left[ d\lambda t+\eta \hat{%
\varepsilon}\lambda x\right] ^{m-1}}{(m-1)!}\hat{\Lambda}_{0}\right\} ,
\notag
\end{eqnarray}%
where the last term is zero for $m=0$. One can easily verify that the
superoperator $\hat{N}_{t}(m)$, Eq. (\ref{nde}), satisfies relation (\ref%
{tt1}).

After lengthy however straightforward calculations we obtain the following
expressions for the initial factorial photocounts moments%
\begin{equation}
\overline{m}_{t}=d\lambda t+\eta \bar{n}\left( 1-\Xi _{1}\right) ,
\label{me}
\end{equation}%
\begin{eqnarray}
\overline{m(m-1)}_{t} &=&\left( d\lambda t\right) ^{2}+2\eta \bar{n}d\lambda
t\left( 1-\Xi _{1}\right) \\
&+&\eta ^{2}\left[ \overline{n(n-1)}\left( 1-\Omega \right) -2\bar{n}\lambda
t\Xi _{2}\right] ,  \notag
\end{eqnarray}%
where%
\begin{equation}
\Xi _{k}\equiv \frac{1}{\bar{n}}\mathrm{Tr}\left[ \frac{\hat{\varepsilon}^{k}%
}{1-\hat{\varepsilon}}\hat{R}_{t}^{0}\rho \right] ,\quad \hat{R}%
_{t}^{0}\equiv \hat{R}_{t}(q=1),  \label{Chik}
\end{equation}%
\begin{equation}
\Omega \equiv \frac{2}{\overline{n(n-1)}}\mathrm{Tr}\left[ \left( \frac{\hat{%
\varepsilon}}{1-\hat{\varepsilon}}\right) ^{2}\hat{R}_{t}^{0}\rho \right] .
\label{Omega}
\end{equation}%
Using Eq. (\ref{www}), the waiting time distribution density is found to be
\begin{eqnarray}
\mathrm{W}_{t}\left( \tau \right) &=&e^{-d\lambda \tau }\left\{ (\lambda
d)^{2}[1-\mathrm{Tr}(\hat{R}_{t}^{0}\rho )]\right.  \notag \\
&+&\left. \mathrm{Tr}[(\hat{J}\hat{R}_{\tau }+\lambda d\hat{\Lambda}_{0}%
\frac{1-\hat{R}_{\tau }}{1-q\hat{\varepsilon}}\hat{\Lambda}_{0})\hat{J}\hat{R%
}_{t}^{0}\rho ]\right\} .  \label{WE}
\end{eqnarray}

In the appendix \ref{apendix1} we show that the dead-time effect can be
incorporated into E-model, however its effect is quite small compared to the
non-unit QE effect, so we disregard it in this paper. In the appendix \ref%
{apendix2} we obtain formulas for Eqs. (\ref{Chik}), (\ref{Omega}) and (\ref%
{WE}) in terms of $\rho_n$ and evaluate them for the coherent, number and
thermal states.

\section{Verifying CPM}

\label{s3}

Basing ourselves on published experimental data \cite{had} we chose the
following numerical values for the model parameters: $\eta =0.6$ for the QE
and $d=5\cdot 10^{-3}$ for the dark counts rate (normalized by the ideal
counting rate). We do not attribute any fixed value to $\lambda $ since our
analysis will be given in terms of the dimensionless $\lambda t$. Many
photodetection quantities in different contexts were reported in, e.g., \cite%
{SD,ueda2,a17,SD1,OMD-JOB,waiting1,waiting4,DMD-JOB05}, so here we shall
consider few of them that could check the validity of either, the SD- or the
E- model in photocounting experiments.
\begin{figure}[t]
\begin{center}
\includegraphics[width=.48\textwidth]{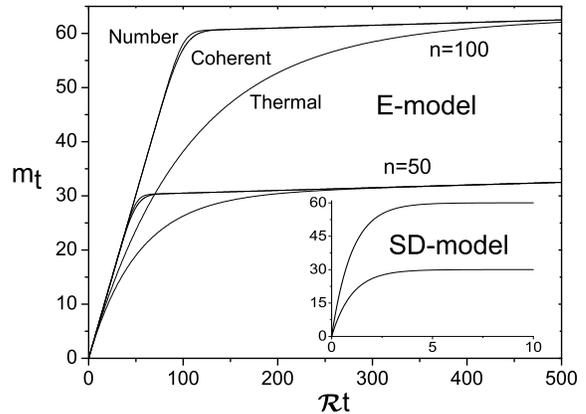} {}
\end{center}
\caption{Mean photocounts number $\bar{m}_{t}$ in E-model for coherent,
number and thermal states (indicated in the figure, the lower curves are
labeled analogously) as function of time for two values of the initial
photon number: the lower curves correspond to $\bar{n}=50$ and the upper --
to $\bar{n}=100$. In the inset we plot $\bar{m}_{t}$ for the SD-model, which
is independent from field state.}
\label{figure1}
\end{figure}

First we analyze the counting statistics. In figure \ref{figure1} we plot $%
\bar{m}_{t}$ as function of $\lambda t$ for both models for two values of
the initial mean photon number, $\bar{n}=50$ and $100$. Initially, $\bar{m}%
_{t}$ increases steeply due to photons absorption, and after some time the
growths turns linear with much smaller slope due to the dark counts. We call
the time interval during which the photons are absorbed (representing the
duration of the steep increase in the number of counts) the \emph{effective
counting time }$t_{E}$. In the E-model $t_{E}$ is proportional to the
initial average photon number, contrary to the SD-model [as seen from the
figure \ref{figure1} and formulas (\ref{msd}) and (\ref{me})]. So the
experimental analysis of the dependence of $t_{E}$ on $\bar{n}$ seems to us
a feasible way for verifying which model could hold in practice, because,
according to the SD-model, $t_{E}$ does not depend on $\bar{n}$. Moreover,
one could also check the validity of each model by verifying whether $\bar{m}%
_{t}$ depends on the initial field state: in the SD-model it is independent
of the field state, while in the E-model $\bar{m}_{t}$ is quite sensible to
it: in figure \ref{figure1} one sees a notable difference between thermal
and coherent states, although not so much between number and coherent
states. This can be explained by a great difference in the values of
Mandel's $Q$-factor \cite{qfactor} characterizing the statistics of photons
in the initial state: it equals $-1$ and $0$ for number and coherent states,
respectively, whereas it is very big ($Q_{th}=\bar{n}$) for the thermal
states with big mean numbers of photons.
\begin{figure}[t]
\begin{center}
\includegraphics[width=.48\textwidth]{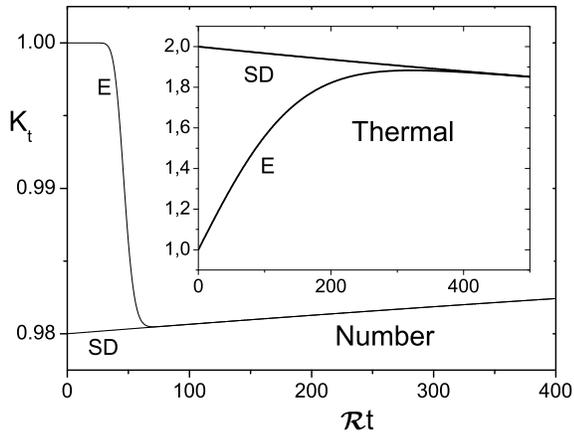} {}
\end{center}
\caption{Normalized second factorial moment $K_{t}$, Eq. (\protect\ref{kkk}%
), for SD- and E- models (as indicated in the graph with abbreviations) for
the number state (and the thermal state in the inset) for $\bar{n}=50$. For
the coherent state one has $K_{t}=1$ at all times for both models.}
\label{figure2}
\end{figure}

Now we analyze the normalized second factorial moment
\begin{equation}
K_{t}\equiv {\overline{m(m-1)}_{t}}/{\overline{m}_{t}^{2}},  \label{kkk}
\end{equation}
for the same initial states with mean photon number $\bar{n}=50$. For the
number and thermal states $K_{t}$ as function of $\lambda t$ is shown in
figure \ref{figure2}, and for the coherent state we get $K_{t}=1$, so it is
not plotted. In the asymptotic time limit and for non-zero dark counts rate,
the same value $K_{\infty }\rightarrow 1$ holds for both models, however the
transient is model dependent. In the SD-model without considering dark
counts $K_{t}$ is time-independent, writing as $K=\overline{n(n-1)}/%
\overline{n}^{2}$ ($\bar{n}$ and $\overline{n(n-1)}$ correspond to the
initial field state), nevertheless it depends on the initial field state: $%
K=2$ for the thermal state and $K=1-1/\bar{n}$ for the number state. By
including the dark counts in the analysis this constant behavior is slowly
modified as time goes on, see figure \ref{figure2}.

In the E-model in the absence of dark counts $K_{t}$ starts at the value
\begin{equation}
\lim_{t\rightarrow 0}K_{t}=\frac{\mathrm{Tr}\left( \hat{\varepsilon}^{2}\rho
\right) }{\left[ \mathrm{Tr}\left( \hat{\varepsilon}\rho \right) \right] ^{2}%
}=\frac{1-\rho _{0}-\rho _{1}}{\left( 1-\rho _{0}\right) ^{2}},
\end{equation}%
which is exactly $1$ for the number state and very close to $1$ for the
thermal state with the chosen values of $\bar{n}$. With the course of time, $%
K_{t}$ attains the same values as for the SD-model (for respective initial
field states) when all the photons have been counted. By taking in account
the dark counts effect such a behavior is slightly modified, yet it is quite
different from the behavior in the SD-model, as shown in the figure \ref%
{figure2}. This is another possible manner for verifying the applicability
of SD- or E- models.

We now turn our attention to the waiting time analysis. It is important to
define the time interval in which we do the average: if one has non-zero
dark counting rate, then by performing the average over a very large time
interval, we shall always get for the mean waiting time the value $\bar{\tau}%
\sim \left( \lambda d\right) ^{-1}$, which is nothing but the mean time
interval between consecutive dark counts. Since experimentally the average
is done over finite time intervals, we shall proceed in the same way: the
mean waiting time for initial times, when the photon number is
significative, is roughly $(\eta \lambda )^{-1}$ (because $\eta \lambda $ is
the effective counting rate), so we shall take the average over a time
interval $\nu =10\,(\eta \lambda )^{-1}$. This means that if one does not
detect consecutive counts within the time $\nu $, such a measurement will
not contribute to the average. In an ideal case this procedure is not
necessary because the probability for registering consecutive clicks
separated by a large time interval is zero.
\begin{figure}[t]
\begin{center}
\includegraphics[width=.48\textwidth]{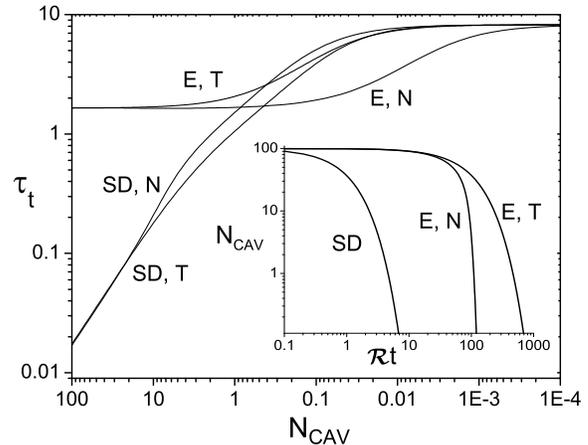} {}
\end{center}
\caption{Mean waiting time $\bar{\protect\tau}_{t}$ as function of $N_{CAV}$
for the number (N) and thermal (T) states for SD- and E- models. While there
are photons in the cavity $\bar{\protect\tau}_{t}$ is constant for the
E-model, but increases with time for the SD-model. In the inset we plot $%
N_{CAV}$ as function of $\protect\lambda t$ for these states (in the
SD-model $N_{CAV}$ is state independent).}
\label{figure3}
\end{figure}

In figure \ref{figure3} we plot the mean waiting time for the SD- and E-
models, for the number and thermal initial states (for the coherent state we
obtain a curve almost identical to the one for a number state) with $\bar{n}%
=100$ as function of the mean photon number in the cavity at the moment of
the first click,
\begin{equation}
N_{CAV}=\mathrm{Tr}\left[ \hat{n}\hat{T}_{t}\rho _{0}\right] =
\left\{
\begin{array}{c}
\bar{n}e^{-\lambda t}\quad \mbox{for SD-model} \\
\bar{n}\Xi _{1} \quad \mbox{for E-model.}%
\end{array}%
\right.  \label{ncav}
\end{equation}%
(For completeness, in the inset of figure \ref{figure3} we plot $N_{CAV}$ as
function of $\lambda t$ for both models.) For the E-model, we see that when $%
N_{CAV}$ becomes less than $1$, the waiting time starts to increase
dramatically due to the dominance of dark counts, which are much more rare
events than absorption of photons. This is a drastic difference from the
ideal case, in which no counts occur after all the photons having been
absorbed, so the mean waiting time saturates at the inverse value of the
counting rate, as shown in \cite{DMD-JOB05}. Moreover, from figure \ref%
{figure3} one verifies that as long as there are photons in the cavity the
mean waiting time is nearly time-independent within the E-model (and truly
independent in the ideal case \cite{DMD-JOB05}), and do increase
substantially in time for SD-model. This is another notable qualitative
difference we suppose one could verify experimentally.

\section{Summary and conclusions}

\label{s4}

In this paper we have generalized the continuous photodetection
model through a careful quantum treatment of non-ideal effects that
are ubiquitous in experiments. We derived general expressions for
the fundamental operations in the presence of non-unit quantum
efficiency and dark counts, and calculated explicitly the
photocounts and the waiting time probability distributions for
initial coherent, number and thermal field states. By calculating
the first and second factorial moments of the photocounts and the
mean waiting time, we showed that in standard photodetection
experiments one could check the applicability of the QJS of SD- or
E- models. Namely, we indicated three different ways for revealing
the actual QJS: (1) quantitatively, one should study the time
dependence of the normalized second factorial photocounts moment.
Qualitatively, we showed that the models can be also distinguished
by measuring: (2) whether the effective detection time depends on
the initial average photon number in the cavity and (3) whether the
mean waiting time is modified as time goes on. To that end we have
considered three different kinds of field in the cavity: the number,
coherent and thermal states; each one on its own permitted to do
comparisons between the two studied QJS's. Results with other kinds
of fields could also be presented here, as for instance the binomial
state or the so-called squeezed state, however, no new physics
related to the goals of the paper appears.  A last remark, if the
experimental data would depart significantly from the theoretical
prediction one should reconsider both models and try to look for
alternative mechanisms to reproduce the outcomes.

In conclusion, we believe that our theoretical treatment could provide clues
for an experimental verification of the CPM, contributing with valuable
insights about the quantum nature of the photodetection in cavities, as well
as giving rise to the possibility of field state manipulation through
detector post-action on the field.

\begin{acknowledgments}
Work supported by FAPESP (SP, Brazil) contract \# %00/15084-5,
04/13705-3. SSM and VVD acknowledge partial financial support from CNPq (DF,
Brazil).
\end{acknowledgments}

\appendix

\section{Cavity damping and dead-time\label{apendix1}}

First we include the effect of cavity damping in our treatment. In
quantum optics experiments the background photons number is
negligible, so we can model the cavity as a thermal reservoir with
zero mean excitations number, described by the standard master
equation \cite{QT1}. Then the UTE equation
in SD-model should be%
\begin{eqnarray}
\dot{\rho} &=&-i\omega \left( \hat{n}\rho -\rho \hat{n}\right) -\frac{%
\lambda }{2}\left( \hat{n}\rho +\rho \hat{n}-2\hat{A}\rho \right)  \notag \\
&&-\frac{\lambda c}{2}\left( \hat{n}\rho +\rho \hat{n}-2\hat{A}\rho \right) ,
\end{eqnarray}%
where $\lambda c$ is the cavity damping rate. From it, following the steps
of sec. \ref{s2} we obtain the no-count superoperator
\begin{equation}
\hat{S}_{t}\rho _{0}=e^{-d\lambda t}\hat{U}_{t}\left( e^{\tilde{q}\tilde{\phi%
}_{t}\hat{A}}\rho _{0}\right) ,\quad \tilde{\phi}_{t}=\frac{1}{p}\left(
1-e^{-\lambda pt}\right) ,  \label{st1}
\end{equation}%
\begin{equation}
p\equiv 1+c,\qquad \tilde{q}\equiv 1-\eta +c=p-\eta .  \label{pq}
\end{equation}%
The value of $c$ should be at the most of order of $10^{-1}$ in order to
make viable the CPM. In this case we see that if one takes into account the
QE drawback, the cavity damping does not modify substantially the resulting
expressions. Therefore we disregard its effect in this paper.

The dead-time effect means that immediately after a click the detector is
unable to register another count within a quite small time interval $x$, $%
\lambda x\ll 1$. In our framework we can describe this effect as the
occurrence of the UTE during the time $x$ immediately after the count, so
the conditioned density operator $\hat{h}$ (\ref{rho_c}) becomes
\begin{eqnarray}
\hat{h}\rho &\equiv &\hat{S}_{t-t_{m}-x}\hat{T}_{x}\hat{J}\hat{S}%
_{t_{m}-t_{m-1}-x}\hat{T}_{x}\hat{J}\cdots \hat{J}\hat{S}_{t_{1}}\rho  \notag
\\
&=&\hat{S}_{t-t_{m}}\hat{\Theta}\hat{J}\hat{S}_{t_{m}-t_{m-1}}\hat{\Theta}%
\hat{J}\cdots \hat{\Theta}\hat{J}\hat{S}_{t_{1}}\rho ,  \label{nd}
\end{eqnarray}%
where the \emph{dead-time superoperator}, under condition $\lambda x\ll 1,$
is found to be
\begin{equation}
\hat{\Theta}\equiv \hat{S}_{-x}\hat{T}_{x}=\exp (x\hat{J})
\end{equation}%
for both SD- and E- models, with respective QJS's.

In SD-model the resulting dead-time superoperator is unbounded, as well as $%
\hat{J}$, so it can bring some mathematical inconsistences. For example, the
$m$-counts superoperator with dead-time effect is found to be
\begin{equation}
\hat{N}_{t}(m)=\hat{S}_{t}\frac{\left[ d\lambda t+e^{d\lambda x}\hat{z}%
/(p\phi _{z})\right] ^{m}}{m!},  \label{nqq12}
\end{equation}%
where
\begin{equation}
\hat{z}\equiv e^{\eta \phi _{x}\hat{A}}-e^{\eta \phi _{x}\hat{A}\exp \left(
-p\lambda t\right) }.
\end{equation}%
If one evaluates, for instance, $\mathrm{Tr}\left[ \sum_{m}m^k\hat{N}%
_{t}(m)\rho \right] $ one will find a divergent result because $\hat{z}$
increases much faster than the decreasing terms.

In the E-model $\hat{J}$ is a bounded superoperator, so the dead-time
corrections will be of order $\eta \lambda x\ll 1$, much less relevant than
the non-unit QE drawback.

\section{ Evaluation of traces}

\label{apendix2}

In this appendix we derive general expressions for both SD- and E- models
and evaluate them for a general initial density operator $\rho =\sum \rho
_{n}|n\rangle \langle n|$ (non-diagonal elements do not contribute to the
trace in the expressions below). We shall analyze three particular field
states: coherent state,
\begin{equation*}
\rho _{n}=e^{-\bar{n}}\bar{n}^{n}/n!\,, \quad \overline{n(n-1)}=\bar{n}^{2},
\end{equation*}
number state,
\begin{equation*}
\rho _{n}=\delta _{n,\bar{n}} \quad \mbox{with integer} \quad \bar{n},
\end{equation*}
and thermal state,
\begin{equation*}
\rho _{n}=(1-\alpha )\alpha^{n}, \quad \alpha =\bar{n}/(\bar{n}+1), \quad
\overline{n(n-1)}=2\bar{n}^{2}.
\end{equation*}

In the SD-model, formula (\ref{fkw}) results in:

\begin{itemize}
\item coherent state%
\begin{equation}
\Phi _{k}^{W}=\bar{n}^{k}\exp \left[ -\eta \bar{n}\phi _{\tau }e^{-\lambda t}%
\right]
\end{equation}

\item number state%
\begin{equation}
\Phi _{k}^{W}=\frac{\bar{n}!}{(\bar{n}-k)!}\left( 1-\eta \phi _{\tau
}e^{-\lambda t}\right) ^{\bar{n}-k}
\end{equation}

\item thermal state%
\begin{equation}
\Phi _{k}^{W}=\frac{k!(1-\alpha )\alpha ^{k}}{\left[ 1-\alpha (1-\eta \phi
_{\tau }e^{-\lambda t})\right] ^{k+1}}.
\end{equation}
\end{itemize}

The formula (\ref{ncav}) yields $N_{CAV}=\bar{n}e^{-\lambda t}$ for all the
states.

In the E-model we need to expand the superoperators as series of $\hat{%
\varepsilon}$ and evaluate the sums. For Eqs. (\ref{Chik}) and (\ref{Omega})
we obtain
\begin{eqnarray}
\Xi _{k} &=&\frac{e^{-\lambda t}}{\bar{n}}\sum_{n,l,m=0}^{\infty }\frac{%
\left( \lambda t\right) ^{m}}{m!}\rho _{n+l+m+k}  \notag \\
&=&\frac{e^{-\lambda t}}{\bar{n}}\sum_{n,m=0}^{\infty }\left( n+1\right)
\frac{\left( \lambda t\right) ^{m}}{m!}\rho _{n+m+k}
\end{eqnarray}%
\begin{eqnarray}
\Omega &=&\frac{2e^{-\lambda t}}{\overline{n(n-1)}}\sum_{n,l,m=0}^{\infty }n%
\frac{\left( \lambda t\right) ^{m}}{m!}\rho _{n+l+m+1}  \notag \\
&=&\frac{e^{-\lambda t}}{\overline{n(n-1)}}\sum_{n,m=0}^{\infty }n(n-1)\frac{%
\left( \lambda t\right) ^{m}}{m!}\rho _{n+m}.
\end{eqnarray}%
Regarding the evaluation of the mean waiting time (\ref{WE}), one needs to
evaluate the expressions
\begin{equation}
\mathrm{Tr}\left[ \Lambda _{0}\left( \frac{\hat{\varepsilon}^{k}}{1-q\hat{%
\varepsilon}}e^{\lambda \hat{\varepsilon}\beta }\rho \right) \Lambda _{0}%
\right] =\Psi _{k}(q,\beta ).
\end{equation}%
\begin{equation}
\mathrm{Tr}[\hat{\varepsilon}^{k}e^{\lambda \hat{\varepsilon}\beta }\rho
]=\Psi _{k}(q=1,\beta ),
\end{equation}%
where%
\begin{equation}
\Psi _{k}(q,\beta )\equiv \sum_{n,l=0}^{\infty }q^{n}\frac{(\lambda \beta
)^{l}}{l!}\rho _{n+l+k}.
\end{equation}

\begin{itemize}
\item For the thermal state we can evaluate the expressions obtained in the
section \ref{subb} directly using the `eigenstate' relation $\hat{\varepsilon%
}\rho =\alpha \rho $ and $\mathrm{Tr}[\Lambda _{0}\rho \Lambda _{0}]=\rho
_{0}$.

\item For the coherent state we use the formula
\[\sum_{k=0}^{\infty }\frac{x^{k}}{k!(k+n)!} =\frac{I_{n}(2\sqrt{x})}{x^{n/2}}, \]
where $I_{k}(x)$ is the
modified Bessel function \cite{prudnikov}, to obtain%
\begin{equation}
\Xi _{k}=\frac{e^{-\lambda t-\bar{n}}}{\bar{n}}\sum_{n=0}^{\infty }(n+1)%
\left(\frac{\bar{n}}{{\lambda
t}}\right)^{(n+k)/2}I_{n+k}(2\sqrt{\bar{n}\lambda t}),
\end{equation}

\begin{equation}
\Omega =\frac{e^{-\lambda
t-\bar{n}}}{\overline{n(n-1)}}\sum_{n=2}^{\infty
}n(n-1)\left(\frac{\bar{n}}{{\lambda
t}}\right)^{n/2}I_{n}(2\sqrt{\bar{n}\lambda t}),
\end{equation}%
\begin{equation}
\Psi _{k}(q,\beta )=e^{-\bar{n}}\left(\frac{\bar{n}}{{\lambda
t}}\right)^{k/2}
\sum_{n=0}^{\infty }\left(\frac{\bar{n}q^{2}}{{\lambda t}}\right)^{n/2}I_{n+k}(2%
\sqrt{\bar{n}\lambda t}),
\end{equation}%
The above series can be transformed in a finite integral using%
\begin{equation*}
\sum_{k=0}^{\infty }t^{k}I_{k+\nu }(z)=\frac{e^{tz/2}}{z^{\nu }}%
\int_{0}^{z}\tau ^{\nu }e^{-t\tau ^{2}/(2z)}I_{\nu -1}(\tau )d\tau ,
\end{equation*}%
valid for $\mathrm{{Re}(\nu )>0.}$

\item For the number state, using $\sum_{k=0}^{n}x^{k}/k!=e^{x}\Gamma
(n+1,x)/n!,$ where $\Gamma (\alpha ,x)=\int_{x}^{\infty }t^{\alpha
-1}e^{-t}dt$ is the incomplete complementary Gamma function
\cite{prudnikov},
we obtain%
\begin{equation}
\Xi _{k}=\frac{\Gamma (\bar{n}-k+2,\lambda t)-\lambda t\Gamma (\bar{n}%
-k+1,\lambda t)}{\bar{n}(\bar{n}-k)!},
\end{equation}%
\begin{equation}
\Omega =\frac{\Gamma (\bar{n}+1,\lambda t)-2\lambda t\Gamma (\bar{n},\lambda
t)+(\lambda t)^{2}\Gamma (\bar{n}-1,\lambda t)}{\overline{n(n-1)}(\bar{n}-2)!%
},
\end{equation}%
\begin{equation}
\Psi _{k}(q,\beta )=q^{\bar{n}-k}e^{\lambda \beta /q}\frac{\Gamma \left(
\bar{n}-k+1,\lambda \beta /q\right) }{(\bar{n}-k)!}.
\end{equation}
\end{itemize}

\end{document}